\documentclass[12pt,showpacs,preprintnumbers,amsmath,amssymb,nofootinbib,superscriptaddress]{article}
\usepackage{graphicx}
\usepackage{dcolumn}
\usepackage{bm}
\usepackage{graphics}
\usepackage{amssymb}
\usepackage{amscd}
\usepackage{afterpage}
\usepackage{float,times}
\usepackage{subfigure}
\usepackage{rotating}
\usepackage{multirow}
\usepackage{fancyheadings}
\usepackage{epsfig}
\usepackage{theorem}
\usepackage{moreverb}
\usepackage{euscript}
\usepackage{psfrag}

\begin{document}
\begin{titlepage}

{\hbox to\hsize{\hfill May 2010 }}

\bigskip \vspace{3\baselineskip}

\begin{center}
{\bf \large 
Stable mass hierarchies and dark matter from hidden sectors in the scale-invariant standard model}

\bigskip

\bigskip

{\bf Robert Foot, Archil Kobakhidze and Raymond R. Volkas \\}

\smallskip

{ \small \it
School of Physics, The University of Melbourne, Victoria 3010, Australia \\
E-mails: rfoot@unimelb.edu.au, archilk@unimelb.edu.au, raymondv@unimelb.edu.au
\\}

\bigskip
 
\bigskip

\bigskip

{\large \bf Abstract}

\end{center}
\noindent 
Scale invariance may be a classical symmetry which is broken
radiatively. This provides a simple way to stabilise the scale of electroweak symmetry
breaking against radiative corrections.  But for such a theory to be fully realistic, it must actually
incorporate a hierarchy of scales, including the Planck and the neutrino mass scales 
in addition to the electroweak scale.  The dark matter sector and the physics responsible for baryogenesis
may or may not require new scales, depending on the scenario.
We develop a generic way of
using hidden sectors to construct a technically-natural hierarchy of scales in the framework
of classically scale-invariant theories. 
We then apply the method to generate the Planck mass and to
solve the neutrino mass and dark matter problems through 
what may be termed the ``scale-invariant standard model''.  
The model is perturbatively
renormalisable for energy scales up to the Planck mass.

\end{titlepage}

\section{Introduction}

It is quite remarkable that almost all the mass of the visible matter in the universe originates from quantum effects that
trigger dimensional transmutation in QCD, even though that theory, at the classical level, 
is strictly invariant under scale transformations when quark masses are neglected. 
The idea that all the elementary particles, including those constituting 
dark matter, obtain their masses through the mechanism of dimensional transmutation is therefore very appealing.
Indeed, a perturbative mechanism for electroweak symmetry breaking in classically scale-invariant models
was presented a long time ago by Coleman and Weinberg \cite{Coleman:1973jx} 
(see also \cite{Gildener:1976ih}).\footnote{Technicolour models of electroweak symmetry
breaking are examples of QCD-like non-perturbative models realising the dimensional transmutation mechanism.} 
Moreover, classical scale invariance can serve as the symmetry that ensures the stability of the electroweak scale
under radiative corrections \cite{Bardeen:1995kv}--\cite{Foot:2007iy}. 

Recently, a number of scale-invariant particle physics models have been 
proposed \cite{Foot:2007as}--\cite{Hempfling:1996ht}. To be fully realistic, any particle physics model, including those
featuring classical scale invariance, must explain neutrino masses, dark matter and the cosmological matter-antimatter
asymmetry.  This often involves the use of other energy scales, not just that of electroweak symmetry breaking.
This presents a particular challenge for classically scale-invariant theories. In fact the above list should then have the
Planck mass scale added to it, because a complete scale-invariant theory must include gravity and a mechanism for
generating its fundamental scale through the quantum scale anomaly.

In this paper we shall discuss scale-invariant models with stable hierarchically-separated mass scales
that are perturbatively renormalisable up to the Planck scale, where the latter feature ensures calculability.
We first describe a simple general formalism using hidden sectors that achieves this purpose.  We then
apply that formalism to generating the Planck scale together with neutrino masses.
We shall associate one of the mass scales with the right-handed neutrino Majorana mass scale, thus implementing the
see-saw mechanism for light neutrino masses in the context of scale-invariant models. 
As already noted, the generation of a radiatively stable mass hierarchy in scale-invariant models is not a 
trivial task.\footnote{In \cite{Meissner:2006zh} a see-saw model for neutrino masses within 
the scale-invariant standard model is discussed. However, the hierarchy between the electroweak 
and see-saw scales in \cite{Meissner:2006zh} is actually based on large-$\log$ radiative 
corrections, thus undermining the whole perturbative approach.  Another attempt was made in \cite{Foot:2007ay},
but the hierarchy there was due to a small Yukawa coupling constant rather than because a genuinely new mass scale
was generated.} 
Because tree-level 
masses are absent in scale-invariant theories, a hierarchy of mass scales can only 
be generated through a hierarchy of dimensionless coupling constants. Such a hierarchy will be technically 
natural --- stable under quantum corrections --- if sectors which contain the different 
mass scales decouple from each other in the limit where the relevant coupling constants vanish \cite{Foot:2007iy}.
Any dark matter sector is ``hidden'' by definition.  We shall demonstrate
that our framework can incorporate mirror dark matter, which actually does not require the generation of a new scale.

\section{Generating a stable hierarchy of scales}

Consider a hidden (high mass scale) sector consisting of a set of scalar fields ($S_1$, $S_2$, $\ldots$) which are 
singlets under the standard model (SM) gauge group. In the limit where these scalars decouple from 
the visible sector involving the SM fields, and, in particular, the standard 
electroweak Higgs doublet field $\phi$, 
the scalar potential separates:
\begin{eqnarray}
V(\phi, S_1, S_2,...) = V(\phi) + V(S_1, S_2, ..).
\end{eqnarray}
In this limit $V(\phi)$ is simply the Coleman-Weinberg potential,
and given the heavy top quark, spontaneous symmetry breaking does not arise.
Thus we have a massless Higgs particle in this limit. However, the $V(S_1, S_2, ..)$
part can induce spontaneous symmetry breaking, leading to $\langle S_j \rangle \neq 0$.
Now, if we allow the hidden sector to couple to the ordinary matter sector via
\begin{eqnarray}
V = \sum_i \lambda_x^i \phi^{\dagger}\phi S_i^2,
\label{ind}
\end{eqnarray}
then the symmetry breaking will be communicated to the electroweak sector. Indeed, for 
some negative $\sum_i\lambda_x^i\langle S_i^2 \rangle$ the interactions in (\ref{ind}) trigger 
nonzero vacuum expectation value (VEV) $\langle \phi^2 \rangle = -\frac{1}{\lambda_{\phi}}\sum_i\lambda_x^i\langle S_i^2 \rangle$, where $\lambda_{\phi}$ 
is the electroweak Higgs self-interaction coupling.  
The hierarchy of scales $\langle \phi \rangle/\langle S_j \rangle$ is thus controlled by 
the adjustable parameters $\lambda_x^i$. The quantum corrections to the light 
mass scale $\langle \phi \rangle$ due to the heavy mass scales $\langle S_j \rangle$ are also entirely controlled by the 
same coupling constants $\lambda_x^i$. In the limit $\lambda_x^i \to 0$ the heavy and light sectors decouple from each other, and $\langle \phi \rangle/\langle S_j \rangle \to 0$. Therefore, no 
radiative correction can significantly disturb the light mass scale $\langle \phi \rangle$, and, hence, we have a technically natural
solution to the hierarchy problem. Technically natural hierarchies can be similarly generated within the hidden sector of $S$-fields as well.  

The simplest case of a hidden sector consisting of just one real scalar $S$
was discussed earlier \cite{Foot:2007iy}. In that case,
$\lambda_x$ induces symmetry breaking in both the SM and hidden sectors. A drawback of this 
scenario is that the Higgs mass must be relatively large, 
$M_H^2 \stackrel{>}{\sim} \sqrt{12} M_t^2$, which means that 
the model does not remain perturbative up to the Planck scale,
and also is not consistent with the constraint from precision electroweak data.
This motivates consideration of the next simplest model consisting
of a Higgs doublet and two real scalars, $S_1$ and $S_2$,
which we consider below.
We shall show that such a model is perturbative up to the Planck scale and consistent with
constraints from precision electroweak data.  The model can also naturally
incorporate neutrino masses via the see-saw mechanism. Dark matter can be introduced by 
extending the hidden sector to include a mirror copy of all the known particles, which we also discuss.

\section{The two-scalar-singlet model}

Let us start by working in the decoupled limit, and just
consider the hidden sector tree-level potential $V_0(S_1, S_2)$,
\begin{eqnarray}
V_0(S_1,S_2) = \frac{\lambda_1}{4} S_1^4 + \frac{\lambda_2}{4} S_2^4 + \frac{\lambda_3}{2} S_1^2 S_2^2,
\label{a1}
\end{eqnarray}
where for simplicity we impose invariance under $S_1 \to - S_1$.  All mass terms and coupling terms other
than the quartic are zero at the classical level because of the imposed scaling symmetry.

Parameterising the fields through
\begin{equation}
S_1 = r\cos\omega, \ S_2 =r\sin\omega,
\label{a2}
\end{equation}
the potential (\ref{a1}) is then rewritten as
\begin{equation}
V_0(r, \omega)=\frac{r^4}{4}\left( \lambda_1\cos^4\omega
+ \lambda_2 \sin^4\omega+ 2\lambda_3 \sin^2\omega\cos^2\omega
\right).
\label{a3}
\end{equation}
This tree-level potential is quantally corrected as per the Coleman-Weinberg
analysis. We shall work in the parameter regime where the one-loop
perturbative correction $\delta V_{1-{\rm loop}}$ is accurate.  Ideally,
one would like to directly minimise the corrected potential $V \simeq V_0 + \delta V_{1-{\rm
loop}}$, but this is impossible to do analytically.  We instead follow the approximate procedure introduced
by Gildener and Weinberg~\cite{Gildener:1976ih} which is valid in our weakly-coupled theory.

The procedure requires us to at first partially ignore the radiative
corrections and to minimise the tree-level potential (\ref{a3}), but with the recognition that in the quantum theory
the parameters $\lambda_{1,2,3}$ become running coupling constants, depending on renormalisation scale $\mu$. 
The tree-level potential has flat radial directions, which means we begin by
taking $\langle r \rangle$ to be arbitrary (but nonzero). 
There are two possibilities, depending on
the sign of $\lambda_3$.  For the case $\lambda_3 > 0$, the possible symmetry breaking patterns are,
\begin{eqnarray}
\sin\omega = 0 \ {\rm with} \ \lambda_1(\Lambda) = 0 \Rightarrow  \langle S_1 \rangle  
                              = \langle r \rangle = v,  \langle S_2 \rangle = 0 \  {\rm or} 
\nonumber \\
\cos\omega = 0 \ {\rm with}  \ \lambda_2(\Lambda) = 0  \Rightarrow  \langle S_2 \rangle
                                = \langle r \rangle = v,  \langle S_1 \rangle = 0. \nonumber \\
\label{eq:solution1}
\end{eqnarray}
Taking the first solution for definiteness, the scale $\Lambda$ is the renormalisation
point where $\lambda_1$ vanishes.  The
dimensionless parameter $\lambda_1$ transmutes into the scale $\Lambda$
in the quantised theory.
This is a manifestation of the scale anomaly of quantum field theory which generates dimensionful
quantities such as masses despite the classical scale invariance.

For the case $\lambda_3 < 0$, both $S_1$ and $S_2$ gain VEVs,
\begin{equation}
\langle S_1 \rangle = \langle r \rangle \left(\frac{1}{1+\epsilon}\right)^{1/2} \equiv v, \qquad 
\langle S_2 \rangle = v \epsilon^{1/2},
\label{eq:solution2}
\end{equation}
where
\begin{equation}
\langle \tan^2\omega \rangle \equiv \epsilon = \sqrt{\frac{\lambda_1(\Lambda)}{\lambda_2(\Lambda)}},
\end{equation}
with
\begin{equation}
\lambda_3(\Lambda)+\sqrt{\lambda_1(\Lambda)\lambda_2(\Lambda)}=0
\label{a66}
\end{equation}
imposed.  As in the previous case, this relation between the Higgs potential parameters
serves to define the renormalisation point, and a dimensionless parameter is transmuted into the scale $\Lambda$. 

The hierarchy between the VEVs of $S_1$ and $S_2$ is determined through the 
parameter $\epsilon$.\footnote{Note that the hierarchy between the VEVs of $S_1$ and $S_2$ is stable under
radiative corrections.  The small parameter $\lambda_1$ receives a 1-loop correction that is proportional to
$\lambda_3^2$, which in turn is equal to $\lambda_1 \lambda_2$ when evaluated at the scale $\Lambda$.  Thus
the correction to $\lambda_1$ is under control provided $\lambda_2$ is not too large, a condition we need in any case to
make our weak-coupling analysis valid.} 
We can thus immediately apply the above to the hierarchy between the Planck mass and the see-saw scale, 
so that 
\begin{equation}
\epsilon\sim \left(M_{\rm see-saw }/M_P \right)^2, 
\end{equation}
where $S_1$ is required to couple to the gravitational scalar 
curvature via ${\cal L} \sim \sqrt{-g} S_1^2 R$,
and its VEV therefore generates the Newton constant \cite{Foot:2007iy}.
The smaller VEV of $S_2$ generates masses for right-handed neutrinos through the Yukawa 
couplings,\footnote{We are not interested in the flavour structure of neutrino mass matrices here,  
so consider flavour-diagonal couplings for simplicity. Note that the discrete $S_1\to -S_1$ 
symmetry prevents couplings of $S_1$ to right-handed neutrinos, but in its absence $S_1$ would also 
contribute to right-handed neutrino masses.}
\begin{equation}
{\cal L}_{\rm Majorana}= \lambda_{M}^{i}\bar \nu_R^i (\nu_R^i)^c S_2 + {\rm H.c.}
\label{bb}
\end{equation}    

We next calculate the tree-level Higgs masses. We first define the
shifted fields
$S_1 = \langle S_1 \rangle + S'_1$, $
S_2 = \langle S_2 \rangle + S_2'$, and
substitute
them into the potential, Eq.~(\ref{a1}). 
Of the two physical scalars only one (call it $S = \sin \omega S_1 - \cos \omega S_2$) gains mass
at tree-level,
\begin{eqnarray}
m_{S}^2 = \lambda_3 v^2 \ \ & {\rm when} & \ \ \lambda_3 > 0 \nonumber \\
m_{S}^2 = 2(\lambda_1 - \lambda_3) v^2 \ \ & {\rm when} & \ \ \lambda_3 < 0.
\label{eq:treemass}
\end{eqnarray}
The other one (call it $s = \cos \omega S_1 + \sin \omega S_2$) remains massless due to a
flat direction in the Higgs potential. It is the pseudo-Goldstone boson (PGB)  of anomalously-broken scale
invariance.

The PGB gains mass from quantal corrections.
The $1$-loop correction along the flat direction in $V_0$
is \cite{Gildener:1976ih}
\begin{equation}
\delta V_{\rm 1-loop}= Ar^4 \ + \ Br^4\log\left(\frac{r^2}{\Lambda^2}\right )~,
\label{14}
\end{equation}
where
\begin{eqnarray}
A=\frac{1}{64\pi^2 \langle r \rangle^4}
\left[3{\rm Tr}\left(M_V^4\log\left(\frac{M_V^2}{\langle r \rangle^2}\right)\right)\right.
\nonumber \\
\left.+{\rm
Tr}\left(M_S^4\log\left(\frac{M_S^2}{\langle r
\rangle^2}\right)\right)-4{\rm
Tr}\left(M_F^4\log\left(\frac{M_F^2}{\langle r \rangle^2}\right)\right)
\right ]~,
\label{15}
\end{eqnarray}
and 
\begin{equation}
B=\frac{1}{64\pi^2 \langle r \rangle^4}\left [3{\rm Tr}M_V^4+{\rm Tr}M_S^4-4{\rm Tr}M_F^4 \right ]~.
\label{16}
\end{equation}
The traces go over all internal degrees of
freedom, with $M_{V,S,F}$ being the tree-level masses respectively for
vectors, real scalars and Dirac fermions evaluated for the given VEV pattern. 
In this simple case of just two scalars, the only masses we need to consider are the scalar
$S$ and the Majorana fermions $\nu_R$ (which means the $4$ multiplying $M_F^4$
in Eq.~(\ref{16}) becomes a $2$).

The extremal condition $\frac{\partial \delta V_{\rm 1-loop}}{\partial
r}|_{r=\langle r \rangle}=0$ tells us that
\begin{equation}
\log\left(\frac{ \langle r \rangle}{\Lambda} \right)=-\frac{1}{4}-\frac{A}{2B}.
\label{17}
\end{equation}
The PGB mass is then \cite{Gildener:1976ih}:
\begin{eqnarray}
m_s^2  & = & \frac{\partial^2 \delta V_{\rm 1-loop}}{\partial r^2 }
|_{r = \langle r \rangle} = 8 B \langle r \rangle^2  \nonumber \\
& = & 
\frac{1}{8\pi^2 \langle r \rangle^2}\left [3{\rm Tr}M_V^4+{\rm Tr}M_S^4-4{\rm Tr}M_F^4 \right ]~.
\label{18}
\end{eqnarray}
Applying this general formula to our theory we obtain,
\begin{eqnarray}
m_s^2 &\simeq & {1 \over 8\pi^2 \langle r \rangle^2} \left[ m^4_{S} - 2\sum m^4_{\nu_R^i} \right]
\label{eq:smass}
\end{eqnarray}

Now, let us make the model realistic by introducing the Higgs doublet $\phi$ into the picture.
Consider the Higgs potential terms,
\begin{eqnarray}
V_0(\phi, S_1, S_2) = \frac{\lambda_\phi}{2} (\phi^{\dagger} \phi)^2 + 
\frac{1}{2} \phi^\dagger\phi (\lambda_{x1} S_1^2
 + \lambda_{x2} S_2^2)~.
\label{eq:VphiS1S2}
\end{eqnarray}
If $\lambda_x < 0$ then $\langle \phi \rangle \neq 0$ is induced, with
\begin{eqnarray}
\langle \phi \rangle^2 = - \frac{\lambda_{x1}}{2\lambda_\phi} \langle S_1 \rangle^2
- \frac{\lambda_{x2}}{2\lambda_\phi}\langle S_2 \rangle^2~,
\end{eqnarray}
where $\langle \phi \rangle=v_{\rm EW}\approx 174$ GeV.\footnote{The solutions in Eqs.\ (\ref{eq:solution1}) 
and (\ref{eq:solution2}) are modified 
due to the additional terms in Eq.\ (\ref{eq:VphiS1S2}). However, because of the hierarchy 
$\langle \phi \rangle/\langle S_i \rangle \lll 1$, these modifications are insignificant and we 
have neglected them.} In the decoupling limit $\lambda_{xi} \to 0$ 
and $\langle \phi \rangle/\langle S_i \rangle \to 0$. The standard see-saw mechanism 
requires the introduction of couplings between left-handed and massive right-handed neutrinos,  
\begin{eqnarray}
{\cal L}_{\rm Dirac} = \lambda_D^i \bar f_L^i \phi \nu_R^i + {\rm H.c.}
\end{eqnarray}
These couplings together with the couplings in (\ref{bb}) generate quantal corrections to
to $\lambda_{x2}$, so that technically
the decoupling limit corresponds to $\lambda_{xi} \to 0$ and $\lambda_M^i \lambda_D^i \to 0$. 

One can obtain a naturalness constraint on the couplings $\lambda_M^i, \lambda_D^i$ by demanding that there
be no fine-tuned cancellation between the tree level and
the 1-loop contribution to $\lambda_{x2}$. This condition implies,
\begin{eqnarray}
\frac{(\lambda_{M}^{i} \lambda_D^i)^2}{16\pi^2} \stackrel{<}{\sim} 
\lambda_\phi \frac{v_{\rm EW}^2}{\langle S_2 \rangle^2}.
\end{eqnarray}
This condition can be rewritten in terms of the physical masses:
\begin{eqnarray}
M^i \stackrel{<}{\sim} \left( \frac{16\pi^2 v_{\rm EW}^4}{m_\nu^i}\right)^{1/3}.
\end{eqnarray}
The most stringent constraint on $M^i$ comes from the most massive neutrino, which from the atmospheric neutrino
anomaly suggests that the lightest $M^i$ satisfies $M^i \stackrel{<}{\sim} 10^7$ GeV. 
A similar bound has been obtained in \cite{Vissani:1997ys}. 

Note that all the couplings in the Higgs potential can be $\lll {\cal O}(1)$, except for $\lambda_{\phi}$. This means that contributions to the renormalization group 
beta-functions from interactions with hierarchically small coupling constants are negligible compared to other relevant contributions steaming from the Standard Model interactions. Namely, the evolution of the electroweak Higgs self-interaction coupling will be governed by $\beta_{\lambda_{\phi}}\simeq \beta_{\lambda_{\phi}}^{\rm SM}$, where $\beta_{\lambda_{\phi}}^{\rm SM}$ is the Standard Model beta-function. This evolution is largely unaffected by the presence of the heavy hidden sectors simply because the interactions between the sectors are assumed to be extremely weak. Therefore, the theory is manifestly perturbative up to the Planck scale, so long
as the Higgs mass is relatively light. Moreover, boundedness of the potential requires 
positivity of $\lambda_{\phi}$ for all energy scales up to the Planck mass, which gives a 
lower bound on the Higgs boson mass. The range of the allowed Higgs boson masses is 
established through the solutions of the corresponding renormalisation group equations. 
In the regime of dominant $\lambda_{\phi}$ coupling the results are similar to those 
obtained within the minimal standard model \cite{Ellis:2009tp}:
 \begin{equation}
 129~{\rm GeV} \lesssim m_h \lesssim 175~{\rm GeV}.
 \label{b1}
 \end{equation}

\section{Incorporating mirror dark matter}

The mirror extension of the standard model \cite{Foot:1991bp} is one of the best motivated models of a hidden sector.
The SM gauge group $G_{SM}$ is extended to $G_{SM} \times G'_{SM}$, where the second factor is isomorphic
to the first but independent of it.  Standard particles are singlets under $G'_{SM}$, while their mirror partners are
singlets under $G_{SM}$.  A discrete parity symmetry that interchanges the two sectors is imposed.  The fact that it is
a parity symmetry means that a left-handed standard fermion is partnered by a right-handed mirror fermion, and so on.
When the see-saw model is extended by a mirror sector, the usual right-handed gauge-singlet neutrino $\nu_R$
is partnered by a left-handed gauge-singlet neutrino $\nu'_L$, where primes denote mirror partners.  They interchange
under mirror parity.  The standard Higgs doublet is partnered by a mirror Higgs doublet.

Mirror models naturally incorporate dark matter candidates, because if a given ordinary atom is stable then so is
its mirror partner. 
It is remarkable that, unlike many popular dark matter
models, the simplest model with unbroken mirror symmetry is capable of 
explaining  \cite{Foot:2008nw, Foot:2010rj} both the DAMA \cite{Bernabei:2008yi} and 
CoGeNT \cite{Aalseth:2010vx} data and also the null results of the other experiments.  
In \cite{Foot:2007as} we considered the minimal scale-invariant 
mirror matter model and demonstrated that radiative electroweak symmetry breaking is possible due to the 
presence of extra bosonic degrees of freedom, in particular the mirror Higgs boson. 
However, the minimal model also requires spontaneous mirror symmetry breaking and is plagued with a 
Landau pole problem at energies much below the Planck mass. In addition, neutrinos are massless in 
the minimal model.   

All these problems can be solved within a simple extension of the minimal model of Ref. \cite{Foot:2007as}. 
It is obtained as a variation of the model considered in the previous section, where now the hidden sector 
is extended to a full mirror sector.  
The scalars $S_1$ and $S_2$ are assigned as singlets of mirror parity, 
that is $S_i\to S_i$. The field $S_2$ couples to the $\nu'_L$'s as well as to the $\nu_R$'s,
\begin{equation}
{\cal L}_{\rm Majorana}= \lambda_{M}^{i}\left[\bar \nu_R^i (\nu_R^i)^c+
\bar \nu^{\prime~i}_L (\nu^{\prime~i}_L)^c\right]S_2 + {\rm H.c.}
\label{ex1}
\end{equation}  
thus generating the see-saw scale in both sectors. 
The scalar potential comprises of (\ref{a1}) and
\begin{eqnarray}
V_0(\phi, \phi^{\prime}, S_i)  & = & \frac{\lambda_{\phi}}{2}\left[ (\phi^{\dagger}\phi)^2
+ (\phi^{\prime \dagger}\phi^{\prime})^2\right]
+ \frac{\lambda_{\phi\phi}}{2}(\phi^{\dagger}\phi)(\phi^{\prime \dagger}\phi^{\prime}) \nonumber\\
& + & \frac{1}{2}( \phi^{\dagger}\phi+\phi^{\prime \dagger}\phi^{\prime})(\lambda_{x1}S_1^2 +\lambda_{x2}S_2^2).
\end{eqnarray}
Besides the VEVs that spontaneously generate the see-saw and Planck 
scales (\ref{eq:solution2})-(\ref{a66})\footnote{Note, that while the mass of heavy scalar $S$ is still given 
by Eq.\ (\ref{eq:treemass}), the mass of lighter scalar $s$ is modified due to the extra contribution 
from right-handed mirror neutrinos. Thus, instead of (\ref{eq:smass}), we have 
now $m_s^2 \simeq  {1 \over 8\pi^2 \langle r \rangle^2} \left[ m^4_{S} - 4\sum m^4_{\nu_R^i} \right]$.}, 
one obtains, for a finite region of parameter space, 
mirror symmetry preserving VEVs that break electroweak and mirror-electroweak gauge invariance:
\begin{equation}
\langle \phi \rangle=\langle \phi^{\prime} \rangle\equiv v^2_{\rm EW} = 
-\frac{\lambda_{x1}+\epsilon \lambda_{x2}}{2\lambda_{\phi}+\lambda_{\phi\phi}}v^2.
\end{equation}
Again, the hierarchy $v^2_{\rm EW}/v^2\sim M^2_{Z}/M^2_{P}\lll 1$ is determined through the hierarchy of dimensionless 
couplings, $-\frac{\lambda_{x1}+\epsilon \lambda_{x2}}{2\lambda_{\phi}+\lambda_{\phi\phi}}\lll 1$. This hierarchy is 
technically natural because of the decoupling limit $\lambda_{xi} \to 0$.

\section{Conclusion}

The mass carried by the visible and dark matter in the universe may have intrinsically quantum origin as implemented 
in classically scale-invariant theories. We have discussed a class of simple scale-invariant models which 
incorporate a stable hierarchy of mass scales, in particular the hierarchy between the Planck, neutrino see-saw  
and the electroweak scales. We have found that the mass of the lightest right-handed neutrino can be as large as 
$\sim 10^7$ GeV, without disturbing the electroweak scale. This naturalness bound on the lightest right-handed 
neutrino suggests that the simplest leptogenesis \cite{Fukugita} scenario is disfavoured due to the insufficient 
CP-violation \cite{Davidson:2002qv}. However, there are other possibilities such as the resonant leptogenesis scenario (see e.g. \cite{Pilaftsis:2003gt}) with nearly degenerate right-handed neutrinos, which remain a viable option. Within the simplest model, perturbativity and stability requirements bound the Higgs boson mass to the range given in Eq.\ (\ref{b1}). Within the same model one can also simultaneously generate the Planck mass. By extending the hidden 
sector to a full mirror sector, we have obtained a viable scale-invariant mirror model with unbroken mirror 
symmetry. In this way one can incorporate dark matter which is in remarkable 
agreement \cite{Foot:2008nw}, \cite{Foot:2010rj} with recent experimental results from the
DAMA \cite{Bernabei:2008yi} and CoGeNT \cite{Aalseth:2010vx} collaborations.

\subsection*{Acknowledgements}

This work was supported in part by the Australian Research Council.

\end{document}